# PHONON RESONATORS : THE BUILDING BLOCKS OF NANOCOMPOSITE ADJUSTING THE ELECTRON – PHONON INTERACTION

V.G.Andreev , L.V.Kravchuk , and S.G.Lebedev ,
Institute for Nuclear Research RAS, 117312, Moscow, Russia

## Abstract

*The nanocrystallite have the finite number of the oscillation modes. Their number increases proportionally to a cube of the characteristic size. Thus the oscillation spectrum of nanocrystal becomes discrete, and the separate modes of oscillations does not interact with each other, that considerably strengthens all phonon modulated processes in a crystal. Covering of such a nanocrystallite with the shielding surface of a material with the higher nuclear weight will allow to create the phonon resonators whose oscillation modes will represent the standing waves and, will be amplified by the resonant manner. The composites made of phonon resonators will allow to produce a perspective functional material for the electronics with adjustable structure and properties. Some new mechanism of HTS based on phonon resonators is proposed.*

The only mode of high-frequency phonon oscillations that can be obtained in metallic devices made with the use of conventional manufactoring processes is that of travelling wave. Under these conditions any perturbation at a point within the crystal lattice will travel with dissipation in all directions from that point and never come back. In a travelling wave mode phonon oscillation fields are random. Hence, the interaction between conduction electrons and the oscillations of the crystal lattice of an electrical conductor has a viscous nature. We propose to replace the travelling wave mode with the standing resonant wave mode for frequencies from 0.1 to 1 *THz* and higher.

The properties of polycrystalline materials substantially depend on their crystallite size [1]. With decreasing of the crystallite size the number of constituent atoms decreases proportionally to a cube of the linear size. With decreasing of atoms number in the crystallite the number of oscillation modes of a crystal lattice also decreases. Nanocrystallite with the size of 2-3 nanometers consist of $N \sim 10^2$ atoms. The full number of oscillation modes in the lattice of such nanocrystallite is *3N*. At such small number the oscillation modes are isolated each from other and do not interact with each other. In that case the oscillation spectrum of separate nanocrystallite is completely harmonic. The composite material consisting from such nonocrystallites, will possess the unique properties. For example, such composite will not be subjected to thermal expansion (due to absence of the unharmonicity), the plasticity and the strength of such alloy will sharply increase as high as 8-10 times, the melting point decreases as compared with the microcrystalline composite. But much more interesting changes, in our opinion, will be possible to expect in the electrodynamics of such nanocrystalline composites. From this point of view the composites made of nanocrystallites which external surface is covered with the thin (in some atomic layer) shielding surface made of element with higher atomic number may be of interest. This gives rise to reflection of a sound wave from a shielding surface and formation of resonant standing wave inside the nanocrystallite. Such "packed" nanocrystallite has been called by the phonon resonant cavity [2]. The upper cutoff frequency of a phonon cavity is determined by the mass of atoms of the nanocrystalline lattice, while the lower cutoff frequency depends upon the screen atoms mass. The screen is transparent in the band between the upper frequency of nanocluster's atoms and the lower frequency of screen's atoms. Hence no standing wave can be excited in this range but only the travelling wave. Standing wave can only be excited in the range below the lower cutoff frequency. Till now the dynamics of oscillations of a lattice has been connected only with the materials based on the travelling wave propagation. As it is well known the resonant oscillations are extremely amplified on a background of other frequencies, therefore all phonon modulated processes will be strengthened.

The structure of the nanocomposite of the present work will function effectively in the cavity, if the phonon mean free path of $h_F$ is large enough so that the phonon is weakly scattered inside the nanocrystallites. I.e. if the condition fulfiled [3]:

$$1/h_F << k_F = \pi \Delta M / (\lambda_F M), \qquad (1)$$

where $k_F$ - the coupling between phonons, incident and reflected from the shell, $\Delta M$ - difference between the atomic (molecular) mass of material and the shell of the crystallite, $\lambda_F$ - phonon wavelength, $M$ - atomic (molecular) mass of the material of the crystallite. If this relation is not satisfied, the phonon scattering will experience earlier than Bragg reflections from the shell.

The choice of $M$ and $\Delta M$ due to the complicated game of the propagation and scattering of phonons in the nanocomposite, as well as their interaction with the electronic subsystem. The ratio $\Delta M / M$ determines the fraction of phonons that have experienced a reflection from the screen. The $1 - \Delta M / M$ nanocrystal phonons leave without experiencing reflection and gets into the nanocrystals following the course of its propagation. Ie value of $\Delta M / M$ regulates the relation between the number of modes of standing and traveling waves. In this sense, we can represent the three basic modes of operation of the nanocomposite phonon resonators:
1  on the traveling waves at $\Delta M / M \to 0$
2  on the standing waves at $\Delta M / M \geq 1$,
3  intermediate regime for $0 \leq \Delta M / M < 1$.

The case 1 is well known, and we will not comment on it. Intermediate mode 3 is interesting due to involvement the traveling as well as the standing waves in the processes of phonon exchange. This case for very small $\Delta M / M \sim$ 3-6% has been considered in [3]. The use of standing waves gives a strengthening of the phonons in individual nanocrystals due to the resonance between the incident and reflected waves. At the same time share $1 - \Delta M / M$ phonons passing in the next crystals without reflection, as divided in the ratio $\Delta M / M / (1 - \Delta M / M)$ on the standing and traveling mode, etc. up to complete absorption of the initial wave in the nanocomposite. The case of small $\Delta M / M$ is good for a pumping systems similar to the photon lasers and is characterized by a relatively small reflection on the screen that leads to a small increase of the resonant modes in the individual nanocrystallites. At the same time, most of the phonons is in the form of a traveling wave, creating the standing resonant modes in the neighboring crystallites along its path up to complete suppression of a traveling wave. It may be that the total number of births of phonons in the case of small $\Delta M / M$ will exceed those for large *ΔM / M* (volume, covered by a traveling wave will depend on its amplitude, the scattering probability and magnitude of *ΔM / M*).

Search of a suitable material for the phonon resonator screen may follow different strategies. We have decided to resort to electrodynamic simulation of this type of activity. For this purpose the cavity is represented by a two-dimensional set of coupled quarter-wave-length resonators (Fig.1), each representing an atom within the crystal lattice [2]. This set of cavities is surrounded at perifery by conducting stubs to prevent outgoing electromagnetic radiation. Each stub is about 1.5 time as long as the cavity. On a microscopic scale these stubs would be analogous to a set of heavy-metal atoms with atomic mass more than that of the nanocrystallite atoms.

With respect to the manufacturing of phonon resonator based devices a question of interest is: how accurately should the screen material be deposited on the nanocrystallite surface? And would any screen defect be acceptable?

Electrodynamical simulation has demonstrated that defects in the form of vacancies (i.e. the absence of one or several of alternate atoms) or atom displacement from node to an interstitial site have almost no effect on the coupling of phonon modes of adjacent cavities. This is favorable for prospective practical applications of phonon resonators.

Besides the pump systems there is the another class of processes, such as superconductivity, where the role of the traveling wave is not so obvious. It is well known the so-called granular superconductors, where the "communication" between the neighboring grains is difficult. These structures consist of superconducting grains separated by thin (several atomic layers) interlayers of normal metal, semiconductor or insulator. A system of such grains is called a Josephson medium. It is noteworthy that if the interlayer between the grains have a thickness of several atomic layers, the Cooper pairs - carriers of the superconducting current can quantum-mechanically tunnel without breaking into individual electrons between neighboring grains of the Josephson medium. Here the interaction of the electron subsystem with phonon modes in single grain turns out on the foreground. What is the source of these phonon modes? Such a source, of course, can be located inside the granules in the form of natural vibrations of atoms and outside it in the form of traveling waves. However, the most probable source - the natural oscillations of atoms in the granules. It is easy to see that this case is analogous to the case of nanocomposite phonon resonators, "working" in the mode of large *ΔM / M*, and perhaps even regime 2 of very large *ΔM / M*. Hovewer the latter case, with high probability, within the above analogy with the Josephson medium is equivalent to the total absence of tunnel junctions between isolated from each other grains. In a such Josephson media the superconductivity can be present in the individual grains, but the global phase coherency in the composite is absent and the current - voltage characteristics of the composite exhibit a resistive nature. Thus, if considered above analogy with the Josephson medium is correct, then the behavior of nanocomposite of phonon resonators *(NPR)* is described by the regime of large *ΔM / M*. In this case, it becomes crucial to ensure the thickness of the membranes - screens at the level of a few atomic layers and the distance between coated grains chosen such as to provide the bulk electrical conductivity of the nanocomposite. In the case of nanostructured josephson media, which is the subject of our works [4-10] where the attention has been payed to anomalous electromagnetics in the nanosized josephson-like structure of granular carbon films. It is believed, the structure of this films consists of the superconductive graphite granules embedded in a matrix of amorphous carbon with the high electroresistance. Thus the sizes of the granules and intergranular layers have a wide scatter that leads to formation of finite superconducting clusters according to percollation mechanism of conductivity in a composite. Production of controllable *NPR* will allow to minimize the spread in the sizes of their granules and layers and to create the nanocomposite superconductor with the josephson link. And there is the possibility to vary the sizes of granules and thickness of layers by the controllable manner and consequently this allow to adjust the value of josephson link between the granules. This gives rise to large variety of possible applications as the logic and memory elements of quantum computers, detectors and generators of the microwave radiation, which can be considered as the essentially new products and prospective objects for intellectual property. Optimization of the structure of the granular *NPR* with the purpose of strengthening of the effects specified above will allow to develop the new extremely prospective functional material for creation of the devices of josephson-like electronics. The activity on creation of *NPR* are lie on a border line of modern nanoelectronics around the world.

The possibility of strengthening of electron's interaction with phonons of definite wave numbers give us some idea of probable gear for phonon medulated high temperature supercobductivity (*HTS*). The phonon mediation so far is considered as one of the key mechanism of *HTS* [11-12]. In the case of phonon mechanism of HTS one can determine some characteristic phonon mode $f_c$ (for example, so called "soft" mode) which is resposible for cooper pairing. Then taking into account the perspective *HTS* core and subsequent screen materials one can develop the phonon resonators with frequency band including $f_c$. This gives rise to strengthening of $f_c$ in the *NPR*. It is believed than as the result the significant strenthening of *HTS* in the *NPR* can be obtained. This expectation is supported by the prediction of *HTS* in small clusters [13-14].

However the covering of nanoclusters with the nanosceens is rather not trivial problem [15-16]. Problems arise due to both the high mobility of nanoclusters - they intensively "stick together" during the deposition from a vapor phase, and also the violent interaction with the atmospheric oxygen. The oxidation of nanocrystallites in an atmosphere can proceeds by the explosive manner due to their extremely developed surface, therefore during deposition from a vapor the measures should be taken on fast and effective quenching of nanoclusters. The molecular layering method (*MLM*) proposed in Russia in the early 50-ies has been shown to be a good approach to solve of such kind of problems [17]. The main idea of the *MLM*, used for precision synthesis of solids of regular structure, consists in the successive building up of structural units of monolayers of a given chemical composition on the surface of the solid matrix. The synthesis is based on chemical reactions between functional groups on the solid surface and molecules of the reagent supplied from outside. At the same time the reagents used and the reaction products should not enter into chemical interaction between them. To gradually build a new layer of material it is necessary to carry out the multiple and alternate (in sequence) processing of the last pairs of related compounds. And every newly formed monolayer of new functional groups must contain active atoms or groups of atoms capable of reacting with a new portion of the same or a different reagent. The experimental data obtained show that the *MLM* may perform atom-by-atom chemical design of nano-surface by repeated alternation of chemical reactions according to a given program.

The subsequent processes of a nanoclusters covering with the nanoscreens and the process of sintering of nanocomposite from a powder of phonon resonant cavities should occur without access of the oxygen. As a result of all listed actions we can obtain the first rather man-made nanocomposite with the controllable properties - the size of grains and the thickness of layers. So far people faced only with the nanocomposites presented by the nature which properties frequently are casual and are not adjustable. It is necessary to expect, that *NPR* will possess an interesting electrodynamic properties. As phonon modes in the nanocrystallites are independent, they cannot interfere and do not attenuate each other, and, therefore it is necessary to expect the strengthening of such phonon attributed processes as the superconductivity. Thus, it is hoped to raise essentially the temperatures of superconducting transition in *NPR*. Till now in the world practice there are no of such kind of works and the influence of dimensional effects on superconductivity is studied only on an example of thin films and nanowires where the effects expected considerably concede as compared with those of *NPR*.

## ACKNOWLEDGEMENTS

The authors would like to thank to Academician of RAS V.A.Matveev for support of this work and we are gratefully acknowledge the financial support from the Russian Foundation of Basic Research through Grant No. 07-08-13516.

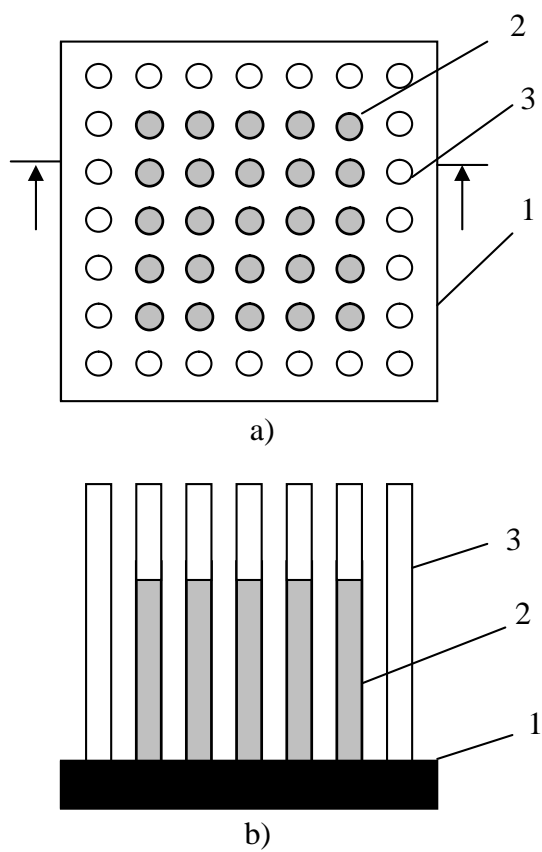

Fig.1. Electrodynamical model of phonon resonator: a) - top view, b) – front view, 1 – mounting conducting plate, 2 – resonators of two-dimensional system, 3 – peripheral stubs in two-dimensional system.